\documentclass[lettersize,journal]{IEEEtran}
\usepackage{array}
\usepackage[caption=false,font=normalsize,labelfont=sf,textfont=sf]{subfig}
\usepackage{textcomp}
\usepackage{stfloats}
\usepackage{url}
\usepackage{algorithm}
\usepackage{titlesec}
\usepackage{verbatim}
\usepackage{graphicx}
\usepackage{float}
\usepackage{cite}
\usepackage{amsmath,amssymb,amsfonts}
\usepackage{xcolor}
\usepackage{algpseudocode}
\usepackage{amsthm}
\usepackage{balance}

\titlespacing{\section}{3pt}{4pt}{2pt}
\titlespacing{\subsection}{2pt}{2pt}{1pt}
\titlespacing{\subsubsection}{1pt}{0pt}{0pt}

\def\BibTeX{{\rm B\kern-.05em{\sc i\kern-.025em b}\kern-.08em
    T\kern-.1667em\lower.7ex\hbox{E}\kern-.125emX}}
\begin{document}

\title{Lyapunov-Based Completion-Aware Scheduling for PDU Set-Based Real-Time XR Traffic}
\author{Baichuan Zhao, Qi Sun, Nan Li and Chih-Lin I,~\IEEEmembership{Life~Fellow,~IEEE}
% \thanks{This work was supported by National Science and Technology Major Project of China under Project GXB-2-2025-5 (\textit{Corresponding author: Qi Sun}).}
\thanks{Baichuan Zhao, Qi Sun (Corresponding author), Nan Li and Chih-Lin I are with China Mobile Research Institute, China (email: \{zhaobaichuan, sunqiyjy, linan, icl\}@chinamobile.com).}
}
\maketitle

\begin{abstract}

Real-time extended reality (XR) services impose stringent throughput, latency, and reliability requirements.
An XR media unit is commonly segmented into a protocol data unit (PDU) set that is useful only when all required PDUs are delivered within its delay budget.
Existing PDU set-aware medium access control (MAC) schedulers generally prioritize transmissions according to urgency or progress, without explicitly accounting for whether a PDU set can still be completed on time.
In this paper, we propose a Lyapunov-based completion-aware MAC scheduler that jointly captures the set-level completion dependency, deadline urgency, and completion feasibility of each head-of-line PDU set.
Long-term quality-of-service (QoS) requirements are represented through virtual debt queues, and resources are allocated according to their marginal contribution to the timely-completion probability of each head-of-line PDU set.
These components are integrated into a drift-plus-penalty scheduling metric that balances long-term reliability requirements against current completion opportunities. 
System-level simulations demonstrate that the proposed scheduler improves PDU set delivery reliability and lower-tail performance under resource contention, while providing robust performance across heterogeneous traffic demands and network conditions.
\end{abstract}

\begin{IEEEkeywords}
PDU set, MAC scheduling, completion-aware, Lyapunov optimization, extended reality
\end{IEEEkeywords}

\section{Introduction}

Extended reality (XR) services impose stringent and heterogeneous requirements on throughput, latency, and reliability. To capture the application-level dependency among protocol data units (PDUs) belonging to the same media unit, 3GPP Release~18 introduced PDU set-based quality of service (QoS) mechanisms, including the PDU set delay budget (PSDB), PDU set error rate (PSER), and PDU set importance information \cite{xr_aware, pdu_set_qos}. These mechanisms enable the base station to schedule PDUs according to PDU set-level delivery requirements rather than treating them independently.

% Extended reality (XR) services impose stringent yet heterogeneous requirements on throughput, latency, and reliability. 
% An application-layer media unit is typically segmented into multiple protocol data units (PDUs) and is useful only when all required PDUs are delivered within a prescribed delay budget. 
% To expose this dependency to the network, 3GPP Release~18 introduced PDU set-based quality-of-service (QoS) mechanisms, including the PDU set delay budget (PSDB), PDU set error rate (PSER), and PDU set importance information \cite{,pdu_set_qos}. 
% These mechanisms allow the base station to schedule PDUs according to their shared application-level semantics rather than treating them independently.

Conventional medium access control (MAC) schedulers, such as proportional fair (PF) and weighted PF, operate mainly at the user or flow level and treat queued data as interchangeable bits \cite{pf_origin,weighted_pf}. 
Consequently, they cannot account for PDU set progress and residual PSDB, failing to capture the all-or-nothing utility of PDU set delivery.

% Conventional medium access control (MAC) schedulers operate mainly at the user or flow level and do not explicitly exploit PDU set information. 
% Proportional fair (PF) scheduling balances instantaneous channel quality and long-term throughput \cite{pf_origin}, while weighted PF further differentiates users or traffic classes through predefined QoS weights \cite{weighted_pf}. 
% However, they both treat queued data as interchangeable bits and therefore cannot account for PDU set progress and residual PSDB, failing to capture the all-or-nothing utility of PDU set delivery.

Recent studies have developed PDU set-aware scheduling algorithms. Delay-aware PF (DAPF) increases the PF priority of PDU sets as their head-of-line (HOL) delay approaches the PSDB~\cite{xr_aware}. 
The Progress-aware scheduler further incorporates the transmitted fraction, favoring PDU sets that are both urgent and close to completion~\cite{progress_aware}. 
The scheduler proposed in~\cite{rate_debt} computes a required service rate from the remaining payload and residual PSDB and tracks the resulting service deficit through a virtual queue. 
Although these methods capture PDU set urgency and partially reflect set-level dependency through transmission progress or service demand, they do not explicitly evaluate whether a PDU set can be completed within its residual PSDB.

Lyapunov optimization offers a principled framework for enforcing long-term constraints through queue-based drift minimization~\cite{neely_delay} and has been widely applied to wireless scheduling~\cite{qos_bs,lyapunov_nr}. 
The scheduler in~\cite{rate_debt} adopts this framework by tracking rate deficits in a virtual queue, but does not directly evaluate the timely-completion probability of the HOL PDU set. 
This limitation motivates a Lyapunov-based completion-aware scheduler that preserves long-term QoS control while explicitly accounting for PDU set completion feasibility.

In this paper, we propose a Lyapunov-based completion-aware MAC scheduler for PDU set-based traffic.
We formulate the scheduling problem as a stochastic optimization problem with long-term success-ratio requirements, and address it using a Lyapunov-based scheduler.
The proposed scheduler maintains a virtual QoS queue for each user and makes scheduling decisions according to both the QoS debt and the completion probability of the HOL PDU set.
The main contributions are summarized as follows.
\begin{itemize}
\item We develop a PDU set-aware MAC scheduling algorithm that jointly accounts for the interdependence among PDUs within each PDU set, the urgency imposed by the residual delay budget, and the feasibility of timely completion under the available service capacity.

\item We derive a Lyapunov-based scheduling metric that combines QoS debt with the marginal improvement in timely-completion probability achieved through resource allocation, thereby balancing long-term reliability requirements against instantaneous completion opportunities.

\item Extensive system-level simulations demonstrate that the proposed scheduler consistently improves PDU set delivery reliability across diverse resource budgets and traffic-load conditions.

\end{itemize}

\textbf{Notations:} 
% $\mathbb{E}[\cdot]$ denotes expectation of random variables. 
$\mathbb{Z}_{\geq 0}$ denotes the set of nonnegative integers.
$\lceil x\rceil$ denotes the smallest integer no smaller than $x$.
% $\mathcal{N}(\mu,\sigma^2)$ denotes the Gaussian distribution with mean $\mu$ and variance $\sigma^2$.
For sequence $X(t)$, denote its long-term time-average expectation as $\bar{X}=\lim_{T\rightarrow\infty}\frac{1}{T} \sum_{t=0}^{T-1}\mathbb{E}[X(t)]$.

\section{System Model and Problem Formulation}

% \subsection{PDU set Traffic Model}

% Each PDU set consists of multiple PDUs associated with the same application-layer media unit and is treated as the basic unit for evaluating timely delivery.

% The main PDU set parameters available for QoS management are summarized as follows:
% \begin{itemize}
% \item \textbf{PDU set Metadata:} Information identifying a PDU set and describing its payload, such as the sequence number and aggregate size.
% \item \textbf{PDU set Delay Budget:} The maximum allowed transfer delay from the arrival of a PDU set at the buffer to its successful delivery.
% \item \textbf{PDU set Error Rate:} The maximum allowed long-term loss ratio of PDU sets, which determines the target PDU set success ratio.
% \item \textbf{PDU set Importance:} Information indicating the relative application-level priority of a PDU set.
% \end{itemize}

% This paper considers the parameters directly relevant to online scheduling, including the PDU set size, remaining payload, residual PSDB, and target success ratio. 
% PDU set importance is not modeled separately but can be reflected through differentiated QoS requirements. 
% Based on these parameters, a completion-aware scheduler should jointly account for the set-level dependency among PDUs, the urgency imposed by the residual PSDB, and the feasibility of timely completion under the available service capacity.

% \subsection{Scheduling Model and Problem Formulation}

We consider a downlink MAC scheduling system with a set of users denoted by $\mathcal{K}$.
Each user maintains one MAC-layer queue, and time is slotted with index $t=0,1,2,\ldots$.
For user $k$, each PDU set contains $B_k$ bits and has a finite PSDB of $D_k$ slots.
The total number of resource blocks (RBs) available for scheduling in each slot is denoted by $R_{\rm tot}$.

Let $A_k(t)$ denote the number of PDU sets arriving at user $k$ in slot $t$.
The corresponding bit arrival is $a_k(t)=A_k(t)B_k$.
Let $Q_k(t)$ denote the queue backlog of user $k$, measured in bits.
The queue contains the bits that have not expired and are still eligible for scheduling.
If a PDU set misses its PSDB, its residual bits are dropped from the queue.
Let $d_k(t)$ denote the amount of data dropped due to PSDB expiration in slot $t$, and the queue evolves as
\begin{equation}
Q_k(t+1)=\max\{Q_k(t)-u_k(t)-d_k(t),0\}+a_k(t),
\label{eq:physical_queue}
\end{equation}
where $u_k(t)$ is the number of successfully delivered bits of user $k$ in slot $t$.
The queue is served in first-in first-out order, and service bits are first applied to the head-of-line (HOL) PDU set.

Let $\alpha_k(t)$ denote the number of RBs allocated to user $k$ in slot $t$.
Let $r_k(t)$ denote the effective per-RB service capability available to the scheduler, which is determined by the current channel condition and transmission configuration.
Then, $u_k(t)$ can be expressed as
\begin{equation}
u_k(t)=\min\{Q_k(t),g(\alpha_k(t),r_k(t))\},
\end{equation}
where the function $g(\cdot)$ maps the allocated RBs and the per-RB service capability to the served data amount, and is monotonically nondecreasing in $\alpha_k(t)$ for a given $r_k(t)$.

PDU set reliability is measured by timely completions.
Let $C_k(t)$ denote the number of PDU sets of user $k$ that are successfully completed in slot $t$ before PSDB expiration.
Let $\eta_k\in[0,1]$ denote the target PDU set success ratio of user $k$, which is determined by its PSER and  reflects PDU set importance.
Then, the PDU set-level QoS requirement is
\begin{equation}
\bar{C}_k\geq \eta_k\bar{A}_k,\quad \forall k\in\mathcal{K},
\label{eq:qos_constraint}
\end{equation}
which means the long-term ratio of timely completions is no smaller than $\eta_k$.

We consider a resource-efficient QoS provisioning problem, where the scheduler minimizes the long-term RB usage subject to the PDU set-level success-ratio constraints, i.e.,
\begin{subequations}
\label{prob:p1}
\begin{IEEEeqnarray}{rl}
\mathbf{P1}:\quad
\min_{\boldsymbol{\alpha}(t)}\  & \bar{f}
\label{prob:p1_objective}\\
\mathrm{s.t.}\ 
& \bar{C}_k\geq \eta_k\bar{A}_k,\  \forall k\in\mathcal{K},
\label{prob:p1_qos}\\
& \sum_{k\in\mathcal{K}}\alpha_k(t)\leq R_{\rm tot},\  \forall t,
\label{prob:p1_resource}\\
& \alpha_k(t)\in\mathbb{Z}_{\geq0},\ 
\forall k\in\mathcal{K},\forall t.
\label{prob:p1_integer}
\end{IEEEeqnarray}
\end{subequations}
where $\bar{f}$ is the long-term average of RB usage $f(t)=\sum_{k\in\mathcal{K}}\alpha_k(t)$, while \eqref{prob:p1_resource} and \eqref{prob:p1_integer} represent per-slot RB budget and nonnegative integer RB allocations, respectively.

\section{Lyapunov-Based PDU set-Aware Scheduling}

In this section, we develop a completion-aware online scheduler for \eqref{prob:p1} using Lyapunov optimization. The long-term PDU set success-ratio constraints are transformed into virtual queue stability conditions, and a probabilistic HOL PDU set completion model is incorporated into the resulting per-slot drift-plus-penalty metric.

% In this section, we derive a completion-aware online scheduler for solving \eqref{prob:p1} based on the Lyapunov optimization framework.
% Specifically, the long-term PDU set success-ratio constraints are transformed into virtual queue stability requirements, based on which a per-slot drift-plus-penalty problem is constructed.
% A probabilistic model of HOL PDU set completion is then incorporated to derive the scheduling metric.

\subsection{Virtual Queue and Lyapunov Drift}

Lyapunov optimization enables online control under long-term average constraints by transforming each constraint into a queue stability requirement.
Accordingly, for the success-ratio constraint in \eqref{eq:qos_constraint}, we introduce a virtual queue $Z_k(t)$ for each user to quantify the accumulated success-ratio debt,
\begin{equation}
Z_k(t+1)=\max\{0,Z_k(t)+\eta_k A_k(t)-C_k(t)\}.
\label{eq:virtual_queue}
\end{equation}
% The virtual arrivals $\eta_kA_k(t)$ represent the required number of timely completions, whereas $C_k(t)$ represents the delivered completions.

If $Z_k(t)$ is stable, then
$ 
\limsup_{T\rightarrow\infty}
\frac{1}{T}\sum_{t=0}^{T-1}
\mathbb{E}[\eta_k A_k(t)-C_k(t)]\leq 0,
$
which implies $\bar{C}_k\geq \eta_k\bar{A}_k$.
Stabilizing $Z_k(t)$ is therefore sufficient to satisfy the long-term QoS constraint.

Problem~\textbf{P1} does not impose an explicit stability constraint on the physical queue $Q_k(t)$.
Under the bounded-arrival and finite-PSDB assumptions, deadline-based dropping prevents expired data from accumulating indefinitely, so the stability of $Q_k(t)$ does not characterize timely-completion QoS.
Nevertheless, $Q_k(t)$ measures the amount of unexpired data awaiting service.
We therefore include it as a backlog regularizer, such that the scheduler also accounts for the buffer pressure caused by useful unserved data.

Let $\boldsymbol{\Theta}(t)=(\mathbf{Z}(t),\mathbf{Q}(t))$ collect the virtual and physical queue states.
To stabilize $Z_k(t)$ while regulating $Q_k(t)$, we define the weighted quadratic Lyapunov function
\begin{equation}
\mathcal{L}(\boldsymbol{\Theta}(t))
=\frac{1}{2}\sum_{k\in\mathcal{K}}
\left[w_ZZ_k^2(t)+w_QQ_k^2(t)\right],
\label{eq:lyapunov_function}
\end{equation}
where $w_Z,w_Q\geq0$ control the contributions of the QoS debt and physical backlog, respectively.
The one-slot drift is
\begin{equation}
\Delta(t)=\mathbb{E}\!\left[
\mathcal{L}(\boldsymbol{\Theta}(t+1))
-\mathcal{L}(\boldsymbol{\Theta}(t)) \mid \boldsymbol{\Theta}(t)\right].
\label{eq:lyapunov_drift}
\end{equation}
Using the queue updates in \eqref{eq:virtual_queue} and \eqref{eq:physical_queue}, together with $(\max\{0,x\})^2\leq x^2$, the drift is upper bounded by
\begin{IEEEeqnarray}{rCl}
\Delta(t)
&\leq& \Gamma
+w_Z\sum_{k\in\mathcal{K}}Z_k(t)
\mathbb{E}\!\left[\eta_kA_k(t)-C_k(t)\right]
\nonumber\\
&&{}+w_Q\sum_{k\in\mathcal{K}}Q_k(t)
\mathbb{E}\!\left[a_k(t)-u_k(t)-d_k(t)\right],
\label{eq:drift_bound}
\end{IEEEeqnarray}
where $\Gamma$ is a finite constant that collects the bounded second-order terms. 
Hereafter, the conditioning on $\boldsymbol{\Theta}(t)$ is omitted for notational simplicity.

To balance queue stability against resource consumption, we add the weighted RB penalty $Vf(t)$ to the drift bound, where $V\geq0$ is an adjustable weight.
Since $A_k(t)$, $a_k(t)$, and $d_k(t)$ are independent of the current scheduling decision, the associated terms can be omitted when minimizing the drift-plus-penalty upper bound, yielding the following per-slot optimization problem:
\begin{subequations}
\label{prob:p2}
\begin{IEEEeqnarray}{rl}
\mathbf{P2}:\quad
\min_{\boldsymbol{\alpha}(t)} \ 
& Vf(t)
{}-w_Z\sum_{k\in\mathcal{K}}Z_k(t)\mathbb{E}[C_k(t)] \nonumber\\
& \quad {}-w_Q\sum_{k\in\mathcal{K}}Q_k(t)\mathbb{E}[u_k(t)] \label{obj:p2}\\
\text{s.t.}\ & \sum_{k\in\mathcal{K}}\alpha_k(t)\leq R_{\rm tot}, \\
& \alpha_k(t)\in\mathbb{Z}_{\geq0},\quad \forall k\in\mathcal{K}.
\end{IEEEeqnarray}
\end{subequations}

\subsection{Service and Completion Probability Estimation}

Solving \eqref{prob:p2} requires estimates of the two expectations in \eqref{obj:p2}, namely, $\mathbb{E}[u_k(t)]$ and $\mathbb{E}[C_k(t)]$.

The expected service can be approximated as
\begin{equation}
\mathbb{E}[u_k(t)]\approx \alpha_k(t)r_k(t),
\label{eq:expected_service}
\end{equation}
which is accurate when the queue backlog is sufficient.

For the term $\mathbb{E}[C_k(t)]$, when at most one PDU set can be completed per user in each slot, it represents the probability that the HOL PDU set is completed in the current slot. Directly maximizing this term, however, overlooks the service opportunities available during the remaining PSDB.

We therefore introduce a look-ahead completion surrogate, denoted by $P_k(\alpha,t)$, which represents the predicted probability that the HOL PDU set is completed within its remaining PSDB when $\alpha$ RBs are allocated in the current slot. 
Unlike $\mathbb{E}[C_k(t)]$, $P_k(\alpha,t)$ corresponds to the eventual timely-completion likelihood, and jointly accounts for the current transmission progress and the remaining service opportunities.

To quantify the effect of $\alpha$ on the completion probability, we model the ratio of the total bits delivered by the PSDB deadline to the PDU set size $B_k$ as a Gaussian distribution, i.e., $X_k(\alpha,t)\sim\mathcal{N}(\rho_k(\alpha,t),\sigma_k^2(t))$.
The HOL PDU set is completed on time if and only if $X_k(\alpha,t)\geq1$. Its completion probability is therefore
\begin{equation}
P_k(\alpha,t)
=\Pr[X_k(\alpha,t)\geq1]
=\Phi\left(\frac{\rho_k(\alpha,t)-1}{\sigma_k(t)}\right),
\label{eq:completion_prob}
\end{equation}
where $\Phi(\cdot)$ denotes the cumulative distribution function (CDF) of the standard Gaussian distribution.

The mean $\rho_k(\alpha,t)$ captures the bits already delivered and expected to be delivered in the current and future slots, i.e.,
\begin{equation}
\rho_k(\alpha,t)=
(\underbrace{B_k-L_k(t)}_{\text{past}}
+
\underbrace{\alpha r_k(t)}_{\text{current}}
+
\underbrace{T_k(t)\bar{\alpha}_k(t)r_k(t)}_{\text{future}})/{B_k}.
\label{eq:rho}
\end{equation}
where $L_k(t)$ and $T_k(t)$ are the remaining payload and the residual PSDB of the HOL PDU set, respectively, and $\bar{\alpha}_k(t)$ is the predicted average number of RBs available to user $k$ per future slot before PSDB expiration.

The standard deviation $\sigma_k(t)$ captures the uncertainty in future service caused by variations in the per-RB service capability.
Let $\sigma_{r,k}^2(t)$ denote the variance of the per-RB service of user $k$.
Assuming that the per-RB service error is independent across future slots, the standard deviation is
\begin{equation}
\sigma_k(t)=
\frac{\bar{\alpha}_k(t)\sqrt{T_k(t)\sigma_{r,k}^2(t)}}{B_k}.
\label{eq:sigma}
\end{equation}

In practice, $r_k(t)$ and $\sigma_{r,k}^2(t)$ can be estimated using the sliding mean and variance of historical per-RB delivery.
In the absence of future scheduling decisions, \(\bar{\alpha}_k(t)\) can be approximated by equally sharing the available RBs among active users, i.e., \(\bar{\alpha}_k(t)=R_{\rm tot}/N_{\rm act}(t)\), where \(N_{\rm act}(t)\) denotes the number of users with nonempty queues.

\subsection{Scheduling Metric}

Approximating $\mathbb{E}[u_k(t)]$ by \eqref{eq:expected_service} and replacing $\mathbb{E}[C_k(t)]$ with the look-ahead surrogate in \eqref{eq:completion_prob} yields a nonlinear integer scheduling problem. 
To enable per-slot scheduling, we adopt a low-complexity greedy heuristic algorithm based on the average gain per RB associated with completing the HOL PDU set of each user.

Specifically, let \(J_k(\alpha,t)\) denote the contribution of user $k$ to the resulting surrogate objective function in \eqref{obj:p2} when $\alpha$ RBs are allocated to it, and is given by
\begin{IEEEeqnarray}{c}
J_k(\alpha,t)
= V\alpha-w_ZZ_k(t)P_k(\alpha,t)-w_QQ_k(t)\alpha r_k(t).
\IEEEeqnarraynumspace\label{eq:local_obj}
\end{IEEEeqnarray}
Let \(m_k(t)\triangleq \min\left(\lceil L_k(t)/r_k(t)\rceil, R_\mathrm{tot}\right)\) denote the number of RBs required to complete the HOL PDU set of user \(k\), capped by the available RB budget.
Thus, the average objective reduction per RB from allocating $m_k(t)$ RBs to user $k$ is
\begin{IEEEeqnarray}{rCl}
M_k(t)
&=& \frac{J_k(0,t)-J_k(m_k(t),t)}{m_k(t)}
\nonumber\\
&=& w_ZZ_k(t)G_k^{\rm cdf}(t)
+w_QQ_k(t)r_k(t)-V,
\label{eq:metric}
\end{IEEEeqnarray}
where
\begin{equation}
G_k^{\rm cdf}(t)
=\frac{P_k(m_k(t),t)-P_k(0,t)}{m_k(t)}
\label{eq:cdf_gain}
\end{equation}
is the average completion-probability gain per RB.

In \eqref{eq:metric}, the first term captures the completion-probability gain weighted by the QoS debt, the second term captures the service gain weighted by the physical queue backlog, and the final term accounts for the per-RB resource cost. 
A larger \(V\) imposes a stronger common penalty on all users, making the scheduler more conservative.
% This mechanism can reduce RB usage while maintaining the required QoS when resources are abundant.
A larger $M_k(t)$ indicates a greater average objective reduction per RB and thus a higher scheduling priority.

To interpret the scheduling behavior induced by the virtual-queue term $w_ZZ_k(t)G_k^{\rm cdf}(t)$, we apply Lagrange's mean value theorem to re-express \eqref{eq:cdf_gain} as the derivative evaluated at an intermediate allocation point.
By the theorem, there exists $\xi_k(t)\in(0,m_k(t))$ such that
\begin{equation}
G_k^{\rm cdf}(t)
=\left.\frac{\partial P_k(\alpha,t)}{\partial \alpha}\right|_{\alpha=\xi_k(t)}.
\end{equation}
Since $P_k(\alpha,t)$ follows a Gaussian CDF and $\partial\rho_k(\alpha,t)/\partial\alpha=r_k(t)/B_k$, the QoS-related term can be expressed as
\begin{IEEEeqnarray}{rCl}
w_ZZ_k(t)G_k^{\rm cdf}(t)
&=& w_ZZ_k(t)\frac{r_k(t)}{B_k}
\frac{1}{\sqrt{2\pi}\sigma_k(t)}
\nonumber\\
& & {}\times
\exp\left(
-\frac{(\rho_k(\xi_k(t),t)-1)^2}{2\sigma_k^2(t)}
\right).
\label{eq:qos_gain_interpretation}
\end{IEEEeqnarray}

Equation~\eqref{eq:qos_gain_interpretation} contains a Gaussian probability density function (PDF) centered at the completion boundary $\rho=1$. When $\rho_k(\xi_k(t),t)$ is far below one, the HOL PDU set is unlikely to be completed on time even if all available RBs in the current time slot are allocated to user $k$. Conversely, when $\rho_k(\xi_k(t),t)$ is far above one, timely completion is already highly likely even without allocating any RB. In either case, the Gaussian PDF takes a small value, thereby reducing the scheduling priority. 
By contrast, when $\rho_k(\xi_k(t),t)$ is close to one, the RB allocation is most likely to alter the completion outcome, resulting in a higher scheduling priority. Therefore, the proposed method accounts for the likelihood that the current RB allocation changes the completion outcome, rather than prioritizing users solely based on urgency or progress. 
Moreover, $Z_k(t)$ and $r_k(t)/B_k$ account for the accumulated QoS debt and the per-RB delivery efficiency, respectively.

\subsection{RB Allocation Rule}

In each slot, active users with positive $M_k(t)$ are sorted in descending order of $M_k(t)$ and sequentially allocated the required $m_k(t)$ RBs until the available RBs are exhausted.
% The complete scheduling procedure is summarized in Algorithm~\ref{alg:lyapunov_scheduler}.

% \begin{algorithm}[htbp]
% \caption{Lyapunov-Based PDU set-Aware Scheduling}
% \label{alg:lyapunov_scheduler}
% \begin{algorithmic}[1]
% \renewcommand{\algorithmicrequire}{\textbf{Input:}}
% \renewcommand{\algorithmicensure}{\textbf{Output:}}
% \Require Queue states $\{Q_k(t),Z_k(t),L_k(t),T_k(t)\}_{k\in\mathcal{K}}$, service statistics $\{r_k(t),\sigma_{r,k}^2(t)\}_{k\in\mathcal{K}}$, and $R_{\rm tot}$
% \Ensure RB allocation $\boldsymbol{\alpha}(t)$
% \State Identify the active-user set $\mathcal{K}_{\rm act}(t)$
% \ForAll{$k\in\mathcal{K}_{\rm act}(t)$}
%     \State Compute $m_k(t)$ and $M_k(t)$ using \eqref{eq:metric}
% \EndFor
% \State Sort active users in descending order of $M_k(t)$
% \State $\boldsymbol{\alpha}(t)\gets\boldsymbol{0}$ and $R_{\rm remain}(t)\gets R_{\rm tot}$
% \ForAll{sorted users $k$}
%     \If{$R_{\rm remain}(t)=0$ or $M_k(t)<0$}
%         \State \textbf{break}
%     \EndIf
%     \State $\alpha_k(t)\gets\min\{m_k(t),R_{\rm remain}(t)\}$
%     \State $R_{\rm remain}(t)\gets R_{\rm remain}(t)-\alpha_k(t)$
% \EndFor
% \end{algorithmic}
% \end{algorithm}

\section{Simulation Results}

\subsection{Simulation Setup}

The proposed scheduler is evaluated using a Sionna-based system-level simulator \cite{sionna}.
We consider the 3GPP indoor-hotspot (InH) deployment, where all XR users are served by one transmission point while other transmission points generate co-channel interference.
The simulation parameters are listed in Table \ref{tab:system_parameters} \cite{rate_debt}.

\begin{table}[htbp]
\caption{System-level simulation parameters}
\vspace{-3mm}
\label{tab:system_parameters}
\centering
\small
\begin{tabular}{@{}>{\raggedright\arraybackslash}p{0.35\columnwidth}
>{\raggedright\arraybackslash}p{0.58\columnwidth}@{}}
\hline
Parameter & Value \\
\hline
Transmission points & 12 ceiling-mounted points \\
Carrier frequency & 3.5 GHz \\
Scheduling bandwidth & 30 RBs, $12 \times 240$ kHz per RB \\
Slot duration & 1 ms \\
Transmit power & 24 dBm (serving), 21 dBm (interfering)\\
Transmission-point antenna array & 64 elements \\
Number of users & 4 VR and 4 CG users\\
User antenna array & Single antenna \\
User velocity & 3 km/h\\
Simulation duration & 10 seconds per realization \\
\hline
\end{tabular}
\end{table}

The XR traffic comprises virtual reality (VR) and cloud gaming (CG) streams, both modeled as periodic PDU set traffic generated according to their respective application frame rates. 
% The two traffic classes differ in PDU set payload size, PSDB, and target success ratio. 
The detailed parameters are summarized in Table~\ref{tab:traffic_parameters}.

\begin{table}[htbp]
\caption{XR traffic and scheduling configuration}
\vspace{-3mm}
\label{tab:traffic_parameters}
\centering
\small
\begin{tabular}{@{}>{\raggedright\arraybackslash}p{0.45\columnwidth}
>{\raggedright\arraybackslash}p{0.50\columnwidth}@{}}
\hline
Parameter & Value \\
\hline
Frame rate & 120 frames/s \\
VR traffic & 45 Mbps, 375 kbits/PDU set \\
VR PSDB and target ratio & 10 ms, $\eta_k=0.98$ \\
CG traffic & 30 Mbps, 250 kbits/PDU set \\
CG PSDB and target ratio & 50 ms, $\eta_k=0.95$ \\
\hline
\end{tabular}
\end{table}

We set $w_Z=1,w_Q = V = 0.01$ to examine the effectiveness of the completion-probability mechanism in the proposed method, and compare it with four baselines, namely PF \cite{pf_origin}, DAPF \cite{xr_aware}, Progress-aware \cite{progress_aware} and Rate Debt \cite{rate_debt}.

% \begin{itemize}
% \item \textbf{PF:} RBs are allocated according to the ratio between the instantaneous service rate and the historical average throughput \cite{pf_origin}.
% \item \textbf{DAPF:} The PF metric is weighted by the normalized HOL delay relative to the corresponding PSDB \cite{}.
% \item \textbf{Progress-aware:} PDU sets with advanced transmission progress and less remaining PSDB are prioritized \cite{progress_aware}.
% \item \textbf{Rate Debt:} A deadline-dependent required service rate is derived, and the resulting service deficit is accumulated in a virtual debt queue for scheduling \cite{rate_debt}.
% \end{itemize}

Scheduler performance is evaluated using the PDU set success ratio of VR and CG streams, defined as the fraction of arrived PDU sets completed within their PSDB. 
Evaluating both traffic classes reveals the reliability tradeoff between delay-sensitive VR and more delay-tolerant CG traffic under a shared RB budget.
All results are averaged over 200 independent simulation runs. 
Unless otherwise specified, the above configuration is used throughout the simulations.

\subsection{Scheduling Behavior in a Three-User Toy Example}

\begin{figure}[htbp]
\centering
\includegraphics[width=0.8\columnwidth]{"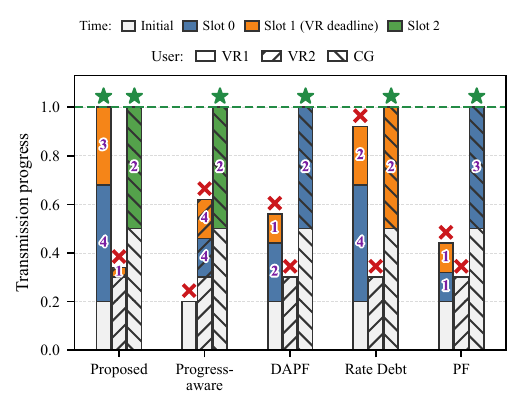"}
\vspace{-6mm}
\caption{Transmission progress in the three-user toy example.
The numbers inside the colored segments indicate the allocated RBs.}
\label{fig:controlled_behavior}
\end{figure}

To illustrate the advantage of the proposed method, we first consider a toy example in which two VR users and one CG user compete for four RBs. 
Figure~\ref{fig:controlled_behavior} compares the scheduling decisions of different methods from a selected critical instant over the subsequent slots. 
For each scheduler, the three bars represent VR1, VR2, and CG, while the vertical axis indicates the normalized transmission progress of their HOL PDU sets. 
The gray segments show the progress accumulated before the selected instant, and the colored segments show the additional progress achieved in successive slots.
At the beginning of the considered slot, the three users exhibit different transmission progress and channel conditions. Their achievable transmission rates satisfy CG $>$ VR1 $>$ VR2. The remaining time before PSDB expiration is 2, 2, and 25 slots for VR1, VR2, and CG, respectively, whereas they require 7, 18, and 2 additional RBs to complete their HOL PDU sets.

As shown in Fig.~\ref{fig:controlled_behavior}, different schedulers make distinct decisions. 
The proposed scheduler concentrates the scarce RBs on the completable VR1 PDU set and completes it by the deadline before serving the less urgent CG user. 
Progress-aware instead prioritizes VR2 because it has made greater progress and has the same residual PSDB as VR1, despite its poor channel condition making timely completion infeasible. 
PF favors the high-rate CG user, while the delay factor in DAPF shifts more resources toward the urgent VR traffic. Rate Debt initially prioritizes VR1 because its high required service rate yields the largest debt-weighted metric. 
After this service reduces the remaining load of VR1, however, the accumulated debt and higher transmission rate of the unserved CG user increase its metric beyond that of VR1 in the second slot. 
Consequently, Rate Debt serves the CG user first and causes VR1 to miss its deadline. 
These decisions leave VR1 with insufficient service under all four baseline schemes. 
This example illustrates that the proposed method jointly captures urgency and completion feasibility under critical resource contention.

\subsection{Reliability and Delay Performance}

\begin{figure}[htbp]
\centering
\includegraphics[width=0.8\columnwidth]{"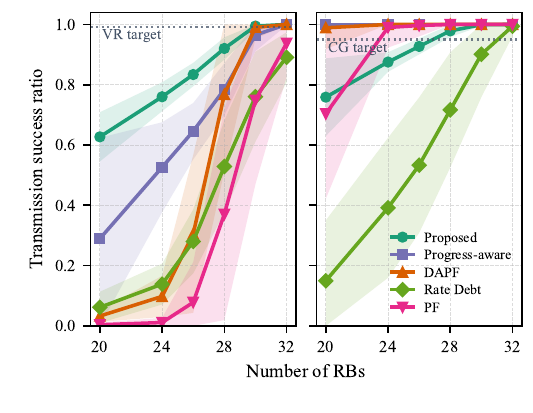"}
\vspace{-6mm}
\caption{Transmission success ratio of PDU sets versus the number of available RBs for VR (left) and CG (right) users.}
\label{fig:rb_sweep}
\end{figure}

Figure~\ref{fig:rb_sweep} compares the PDU set transmission success ratios as the RB budget increases from the resource-limited to the resource-sufficient regime. 
The proposed scheduler achieves the highest VR success ratio with the smallest variation and reaches the VR reliability target at around $R_\mathrm{tot}=30$. 
Under limited resources, Progress-aware and DAPF tend to serve the less delay-sensitive CG users before sufficient resources are preserved for VR traffic.
In contrast, the proposed scheduler improves VR reliability while maintaining CG reliability at a comparatively high level, thereby achieving a better balance between the two traffic classes.

\begin{figure}[htbp]
\centering
\includegraphics[width=0.8\columnwidth]{"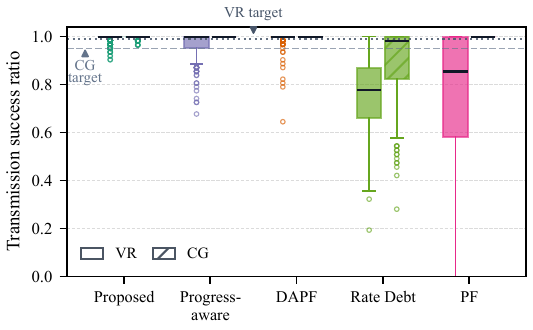"}
\vspace{-6mm}
\caption{Distribution of the per-user PDU set success ratio at $R_\mathrm{tot}=30$.}
\label{fig:success_distribution}
\end{figure}

Figure~\ref{fig:success_distribution} further compares the distributions of per-user PDU set success ratios for the case of $R_\mathrm{tot}=30$, at which the system can approximately support the traffic load.
The proposed method produces the most concentrated VR success-ratio distribution around the target, with the shortest lower tail. 
In comparison, DAPF and Progress-aware exhibit occasional low-reliability outcomes, while Rate Debt and PF show substantially broader distributions extending toward lower success ratios.
The CG distributions further show that the improved robustness of delay-sensitive VR traffic is achieved while maintaining the target of the more delay-tolerant CG traffic.

\begin{figure}[htbp]
\centering
\includegraphics[width=0.8\columnwidth]{"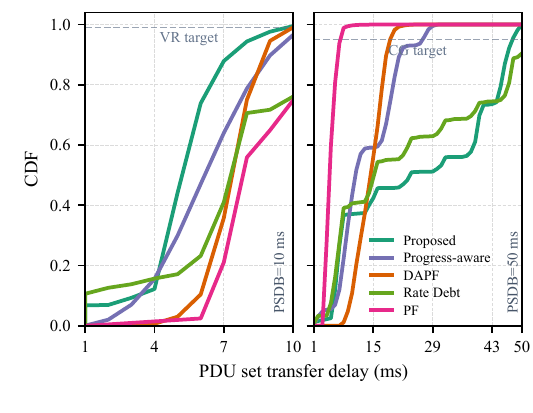"}
\vspace{-6mm}
\caption{PDU set transfer-delay CDFs for (top) VR and (bottom) CG traffic, $R_\mathrm{tot}=30$.}
\label{fig:delay_cdf}
\end{figure}

Figure~\ref{fig:delay_cdf} shows the CDFs of PDU set transfer delays, where the terminal CDF value represents the fraction of PDU sets successfully delivered. 
For VR traffic, the proposed scheduler achieves both a rapid initial increase in the CDF and the highest terminal value, indicating low transfer delays for successfully delivered PDU sets together with the highest delivery reliability. 
Progress-aware provides comparable delays among successful deliveries but has a lower terminal value due to more frequent drops. 
For CG traffic, several baseline schedulers achieve shorter transfer delays by allocating more resources. 
In contrast, the proposed scheduler postpones part of the CG service to protect urgent VR traffic, while still meeting the CG reliability target.

\begin{figure}[htbp]
\centering
\includegraphics[width=0.8\columnwidth]{"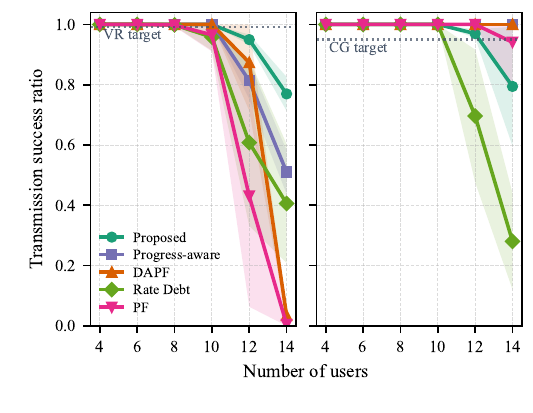"}
\vspace{-6mm}
\caption{PDU set transmission success ratio for VR (left) and CG (right) traffic versus the total number of users, $R_\mathrm{tot}=40$.}
\label{fig:user_sweep}
\end{figure}

To evaluate scheduler scalability, Fig.~\ref{fig:user_sweep} shows the performance as the number of users increases under a more generous RB budget of $R_{\rm tot}=40$, with equal numbers of VR and CG users.
All schedulers achieve nearly lossless transmission with up to eight users.
As the load increases, the proposed scheduler maintains the highest average VR reliability with low variability.
At 14 users, Progress-aware and DAPF preserve CG reliability at the expense of VR traffic, whereas the proposed scheduler provides a more balanced reliability degradation across the two traffic classes.

\section{Conclusion}

In this paper, we proposed a Lyapunov-based completion-aware MAC scheduler for PDU set-based real-time XR traffic with heterogeneous delay and reliability requirements. 
The proposed scheduler jointly accounts for the set-level completion dependency, deadline urgency, and completion feasibility of each head-of-line PDU set. 
By transforming long-term PDU set success-ratio requirements into virtual QoS debt queues and quantifying the improvement in timely-completion probability induced by resource allocation, the proposed method balances long-term reliability requirements against current completion opportunities. 
System-level simulations demonstrated that the proposed scheduler improves PDU set delivery reliability and lower-tail performance under resource contention, while maintaining robust performance across heterogeneous traffic demands and network conditions.

\bibliographystyle{IEEEtran}
\bibliography{references}

\end{document}